\documentstyle[twoside]{article}

\catcode`\@=11
\long\def\@makefntext#1{
\protect\noindent \hbox to 3.2pt {\hskip-.9pt  
$^{{\eightrm\@thefnmark}}$\hfil}#1\hfill}               

\def\thefootnote{\fnsymbol{footnote}}
\def\@makefnmark{\hbox to 0pt{$^{\@thefnmark}$\hss}}    
        
\def\ps@myheadings{\let\@mkboth\@gobbletwo
\def\@oddhead{\hbox{}
\rightmark\hfil\eightrm\thepage}   
\def\@oddfoot{}\def\@evenhead{\eightrm\thepage\hfil
\leftmark\hbox{}}\def\@evenfoot{}
\def\sectionmark##1{}\def\subsectionmark##1{}}

\oddsidemargin=\evensidemargin
\renewcommand{\thefootnote}{\fnsymbol{footnote}}

\newcounter{sectionc}\newcounter{subsectionc}\newcounter{subsubsectionc}
\renewcommand{\section}[1] {\vspace{12pt}\addtocounter{sectionc}{1} 
\setcounter{subsectionc}{0}\setcounter{subsubsectionc}{0}\noindent 
        {\tenbf\thesectionc. #1}\par\vspace{5pt}}
\renewcommand{\subsection}[1] {\vspace{12pt}\addtocounter{subsectionc}{1} 
        \setcounter{subsubsectionc}{0}\noindent 
        {\bf\thesectionc.\thesubsectionc. {\kern1pt \bfit #1}}\par\vspace{5pt}}
\renewcommand{\subsubsection}[1] {\vspace{12pt}\addtocounter{subsubsectionc}{1}
        \noindent{\tenrm\thesectionc.\thesubsectionc.\thesubsubsectionc.
        {\kern1pt \tenit #1}}\par\vspace{5pt}}
\newcommand{\nonumsection}[1] {\vspace{12pt}\noindent{\tenbf #1}
        \par\vspace{5pt}}

\newcounter{appendixc}
\newcounter{subappendixc}[appendixc]
\newcounter{subsubappendixc}[subappendixc]
\renewcommand{\thesubappendixc}{\Alph{appendixc}.\arabic{subappendixc}}
\renewcommand{\thesubsubappendixc}
        {\Alph{appendixc}.\arabic{subappendixc}.\arabic{subsubappendixc}}

\renewcommand{\appendix}[1] {\vspace{12pt}
        \refstepcounter{appendixc}
        \setcounter{figure}{0}
        \setcounter{table}{0}
        \setcounter{lemma}{0}
        \setcounter{theorem}{0}
        \setcounter{corollary}{0}
        \setcounter{definition}{0}
        \setcounter{equation}{0}
        \renewcommand{\thefigure}{\Alph{appendixc}.\arabic{figure}}
        \renewcommand{\thetable}{\Alph{appendixc}.\arabic{table}}
        \renewcommand{\theappendixc}{\Alph{appendixc}}
        \renewcommand{\thelemma}{\Alph{appendixc}.\arabic{lemma}}
        \renewcommand{\thetheorem}{\Alph{appendixc}.\arabic{theorem}}
        \renewcommand{\thedefinition}{\Alph{appendixc}.\arabic{definition}}
        \renewcommand{\thecorollary}{\Alph{appendixc}.\arabic{corollary}}
        \renewcommand{\theequation}{\Alph{appendixc}.\arabic{equation}}
        \noindent{\tenbf Appendix \theappendixc #1}\par\vspace{5pt}}
\newcommand{\subappendix}[1] {\vspace{12pt}
        \refstepcounter{subappendixc}
        \noindent{\bf Appendix \thesubappendixc. {\kern1pt \bfit #1}}
        \par\vspace{5pt}}
\newcommand{\subsubappendix}[1] {\vspace{12pt}
        \refstepcounter{subsubappendixc}
        \noindent{\rm Appendix \thesubsubappendixc. {\kern1pt \tenit #1}}
        \par\vspace{5pt}}

\newcommand{\textlineskip}{\baselineskip=13pt}
\newcommand{\smalllineskip}{\baselineskip=10pt}

\def\eightcirc{
\begin{picture}(0,0)
\put(4.4,1.8){\circle{6.5}}
\end{picture}}
\def\eightcopyright{\eightcirc\kern2.7pt\hbox{\eightrm c}} 

\newcommand{\copyrightheading}[1]
	{\vspace*{-2.5cm}\smalllineskip{\begin{flushright}
	WM-00-111 \end{flushright}
         }}

\def\abstracts#1#2#3{{
        \centering{\begin{minipage}{4.5in}\baselineskip=10pt\footnotesize
        \parindent=0pt #1\par 
        \parindent=15pt #2\par
        \parindent=15pt #3
        \end{minipage}}\par}} 

\newcommand{\bibit}{\nineit}

\renewenvironment{thebibliography}[1]
        {\frenchspacing
         \ninerm\baselineskip=11pt
         \begin{list}{\arabic{enumi}.}
        {\usecounter{enumi}\setlength{\parsep}{0pt}
         \setlength{\leftmargin 12.7pt}{\rightmargin 0pt} 
         \setlength{\itemsep}{0pt} \settowidth
        {\labelwidth}{#1.}\sloppy}}{\end{list}}

\newcounter{itemlistc}
\newcounter{romanlistc}
\newcounter{alphlistc}
\newcounter{arabiclistc}

\newcommand{\fcaption}[1]{
        \refstepcounter{figure}
        \setbox\@tempboxa = \hbox{\footnotesize Fig.~\thefigure. #1}
        \ifdim \wd\@tempboxa > 5in
           {\begin{center}
        \parbox{5in}{\footnotesize\smalllineskip Fig.~\thefigure. #1}
            \end{center}}
        \else
             {\begin{center}
             {\footnotesize Fig.~\thefigure. #1}
              \end{center}}
        \fi}

\newcommand{\tcaption}[1]{
        \refstepcounter{table}

        \setbox\@tempboxa = \hbox{\footnotesize Table~\thetable. #1}
        \ifdim \wd\@tempboxa > 5in
           {\begin{center}
        \parbox{5in}{\footnotesize\smalllineskip Table~\thetable. #1}
            \end{center}}
        \else
             {\begin{center}
             {\footnotesize Table~\thetable. #1}
              \end{center}}
        \fi}

\def\@citex[#1]#2{\if@filesw\immediate\write\@auxout
        {\string\citation{#2}}\fi
\def\@citea{}\@cite{\@for\@citeb:=#2\do
        {\@citea\def\@citea{,}\@ifundefined
        {b@\@citeb}{{\bf ?}\@warning
        {Citation `\@citeb' on page \thepage \space undefined}}
        {\csname b@\@citeb\endcsname}}}{#1}}

\newif\if@cghi
\def\cite{\@cghitrue\@ifnextchar [{\@tempswatrue
        \@citex}{\@tempswafalse\@citex[]}}
\def\citelow{\@cghifalse\@ifnextchar [{\@tempswatrue

        \@citex}{\@tempswafalse\@citex[]}}
\def\@cite#1#2{{$\null^{#1}$\if@tempswa\typeout
        {IJCGA warning: optional citation argument 
        ignored: `#2'} \fi}}

\def\pmb#1{\setbox0=\hbox{#1}
        \kern-.025em\copy0\kern-\wd0
        \kern.05em\copy0\kern-\wd0
        \kern-.025em\raise.0433em\box0}

\def\fnt#1#2{\footnotetext{\kern-.3em
        {$^{\mbox{\scriptsize #1}}$}{#2}}}

\def\fpage#1{\begingroup
\voffset=.3in
\thispagestyle{empty}\begin{table}[b]\centerline{\footnotesize #1}
        \end{table}\endgroup}

\def\runninghead#1#2{\pagestyle{myheadings}
\markboth{{\protect\footnotesize\it{\quad #1}}\hfill}
{\hfill{\protect\footnotesize\it{#2\quad}}}}
\font\tenrm=cmr10
\font\tenit=cmti10 
\font\tenbf=cmbx10
\font\bfit=cmbxti10 at 10pt
\font\ninerm=cmr9
\font\nineit=cmti9

\font\eightrm=cmr8

\def\qed{\hbox{${\vcenter{\vbox{                        
   \hrule height 0.4pt\hbox{\vrule width 0.4pt height 6pt
   \kern5pt\vrule width 0.4pt}\hrule height 0.4pt}}}$}}

\renewcommand{\thefootnote}{\fnsymbol{footnote}}        

\begin{document}

\runninghead{A Lot of Flavor Physics from a Little Symmetry}{A Lot of Flavor
Physics from a Little Symmetry}

\normalsize\textlineskip
\thispagestyle{empty}
\setcounter{page}{1}

\copyrightheading{}                     

\vspace*{0.81truein}

\fpage{1}

\centerline{\bf A LOT OF FLAVOR PHYSICS FROM A LITTLE
SYMMETRY\footnote{Talk presented by R.F. Lebed at DPF 2000, Ohio State
Univ., Columbus, OH, 9--12 Aug.\ 2000.}}
\vspace*{0.37truein}
\centerline{\footnotesize ALFREDO ARANDA {\it and\/} CHRISTOPHER D. CARONE}
\vspace*{0.015truein}
\centerline{\footnotesize\it Nuclear and Particle Theory Group,
College of William and Mary}
\baselineskip=10pt
\centerline{\footnotesize\it Williamsburg, Virginia 23187-8795, USA}
\vspace*{10pt}
\centerline{\footnotesize RICHARD F. LEBED}
\vspace*{0.015truein}
\centerline{\footnotesize\it Department of Physics and Astronomy,
Arizona State University}
\baselineskip=10pt
\centerline{\footnotesize\it Tempe, Arizona 85287-1504, USA}
\vspace*{0.225truein}
\centerline{\footnotesize October 2000}

\vspace*{0.21truein}
\abstracts{Recent neutrino parameter measurements place increasingly
stringent constraints on acceptable supersymmetric theories of flavor.
A very fruitful approach is the application of non-Abelian discrete
gauged flavor symmetries $G_f$. We discuss a highly successful model
using $G_f = T^\prime \times Z_3$, where $T^\prime$ is a non-Abelian
group based on the symmetries of a tetrahedron.  This model reproduces
the basic features of the U(2) model, and also predicts neutrino
masses and mixing angles consistent with either the SMA or LMA
scenarios.}{}{}
\textlineskip                   
\vspace*{12pt}                  

\textheight=7.8truein
\setcounter{footnote}{0}
\renewcommand{\thefootnote}{\alph{footnote}}

\noindent
The family replication problem and the strongly hierarchical pattern
of fermion masses and mixing angles has remained one of the more
intractable problems in particle physics.  Any acceptable flavor
theory must explain why, for example, $m_t/m_e \approx$ 350,000 while
$|V_{ub}| \approx 3 \times 10^{-3}$.  The situation has become even
more interesting with the observation of nonzero neutrino masses and
mixings\cite{neut} that do not appear to follow the same hierarchies
as in the charged fermion sector.  Indeed, the combined data of
existing experiments (omitting LSND\cite{LSND}) suggests a large
mixing angle due to the atmospheric $\nu$ deficit,
\begin{equation}
\sin^2 2\theta_{23} > 0.8, \hspace{2em} 10^{-3} {\rm eV}^2 \le \Delta
m_{23}^2 \le 10^{-2} {\rm eV}^2 ,
\end{equation}
and a solar $\nu$ deficit explained by either the small mixing angle
(SMA) MSW solution
\begin{equation}
2 \times 10^{-3} \le \sin^2 2\theta_{12} \le 10^{-2}, \hspace{2em} 4
\times 10^{-6} {\rm eV}^2 \le \Delta m_{12}^2 \le 10^{-5} {\rm eV}^2 ,
\end{equation}
or the (now somewhat more favored) large mixing angle (LMA) MSW
solution
\begin{equation}
\sin^2 2\theta_{12} > 0.7, \hspace{2em} 6 \times 10^{-6} {\rm eV}^2
\le \Delta m_{12}^2 \le 5 \times 10^{-5} {\rm eV}^2 .
\end{equation}
with three active neutrino species.

We consider the possibility that a spontaneously-broken horizontal
symmetry $G_f$ may explain the totality of fermion mass and mixing
angle data.  We work in the context of supersymmetry (SUSY), which
renders the hierarchy between the scale of flavor physics and the weak
scale stable against radiative corrections.  Flavor-changing neutral
current (FCNC) effects, which are present in generic SUSY models, can
be brought under control by the same symmetry $G_f$ if it leads to a
near mass degeneracy between the first two generations of squarks or
sleptons in the low-energy effective theory.  We require this to be
the case.

We advocate the study of discrete gauge symmetries for $G_f$.  It has
long been argued\cite{global} that global symmetries are violated by
quantum gravity effects and may therefore be inconsistent as
fundamental symmetries of nature.  Alternately, continuous gauge
symmetries in the context of SUSY tend to give rise to excessive
FCNC's through D-term interactions.\cite{KMY} Discrete gauge
symmetries avoid both of these pitfalls.  Theories with global and
gauged discrete symmetries differ in that there is information
contained in topological defects found only in the
latter.\cite{KW,Banks} For model-building purposes, the relevant
low-energy constraint for discrete gauge symmetries is an anomaly
cancellation condition linear in $G_f$ and quadratic in non-Abelian
Standard Model gauge groups.\cite{anom} Our model satisfies this
condition.

The central feature of discrete groups is a finite number of
inequivalent irreducible matrix representations (reps). Discrete
groups tend to have a rich representation structure, which opens many
possibilities for assigning quantum numbers in model building.  Reps
of dimension $>$1 appear only in non-Abelian groups.  In particular,
doublets are used to impose the near-degeneracy of sparticle masses in
the first two generations, while the large value of $m_t$ suggests
that the third generation is distinguished: a ${\bf 2} \oplus {\bf 1}$
structure.

A model\cite{U2} based on the continuous group $G_f = {\rm U}(2)$ is
particularly successful in reproducing all observed charged fermion
masses and mixings, especially when combined with an SU(5) GUT.  The
Yukawa matrix textures are given by
\begin{equation}
Y_D \sim Y_L \sim \left( \begin{array}{ccc} 0 & \epsilon^\prime & 0 \\
-\epsilon^\prime & \epsilon & \epsilon \\ 0 & \epsilon & 1 \end{array}
\right) , \hspace{2em} Y_U \sim \left( \begin{array}{ccc} 0 &
\epsilon^\prime \epsilon & 0 \\ -\epsilon^\prime \epsilon & \epsilon^2
& \epsilon \\ 0 & \epsilon & 1 \end{array} \right) ,
\end{equation}
where the symmetry is broken at two scales, $U(2)
\stackrel{\epsilon}{\longrightarrow} U(1)
\stackrel{\epsilon^\prime}{\longrightarrow} {\rm nothing},$ with
$\epsilon \approx 0.02$ and $\epsilon^\prime \approx 0.004$.  In order
to produce these textures, the U(2) model requires {\bf 1}, {\bf 2},
and {\bf 3} reps and the multiplication rule ${\bf 2} \otimes {\bf 2}
= {\bf 3} \oplus {\bf 1}$.

One finds\cite{PLB} that the smallest group satisfying the same
conditions is $T^\prime$, the group of proper rotations leaving a
regular tetrahedron invariant in the SU(2) double covering of SO(3).
Moreover, including an extra $Z_3$ factor, so that $G_f = T^\prime
\times Z_3$, is the minimal extension needed to reproduce U(2) model
textures and satisfy discrete anomaly cancellation conditions.
Additionally, the rich representation structure of $T^\prime$ allows
for neutrinos to be placed in different reps than the charged
fermions, which is the origin of different hierarchies in the two
sectors.  The symmetry breaking is $T^\prime \times Z_3
\stackrel{\epsilon}{\longrightarrow} Z_3^{\rm diag}
\stackrel{\epsilon^\prime}{\longrightarrow} {\rm nothing},$ and with
the charge assignments presented in \cite{PLB}, the light neutrino
mass matrix (via the seesaw mechanism) is
\begin{equation}
M_{LL} \sim \left( \begin{array}{ccc} (\epsilon^\prime/\epsilon)^2 &
\epsilon^\prime/\epsilon & \epsilon^\prime/\epsilon \\
\epsilon^\prime/\epsilon & 1 & 1 \\ \epsilon^\prime/\epsilon & 1 & 1
\end{array} \right) .
\end{equation}
The 23 mixing angle is clearly $O(1)$; since $\epsilon^\prime/\epsilon
= O(1/5)$ is neither especially large nor small, adjustment of the
implicit $O(1)$ coefficients characteristic of an effective field
theory allows for either LMA or SMA solutions for $\theta_{12}$.

We present in Table~1 a fit to the observable neutrino mass and mixing
parameters for LMA; the full fit also includes the charged fermion
parameters, but is excellent and hence omitted for brevity.  A
similarly excellent SMA fit appears in \cite{PRD}.  The unknown $O(1)$
coefficients $c_i$ are constrained to lie approximately in the range
$1/r \le |c_i| \le r$ by adding terms to the $\chi^2$ function of form
$\Delta \chi_i^2 = ( \ln |c_i| / \ln r )^2$; for this fit, $r=3$.  One
finds $\chi^2_{\rm min} = 11.8$, smaller than the total number of
fermion observables (16), which we argue\cite{PRD} is the figure of
merit.

\begin{table}
\tcaption{Data and fit values for the LMA scenario in the $T^\prime
\times Z_3$ model ($\nu$ sector).}
\centerline{\footnotesize\smalllineskip
\begin{tabular}{lll}\\
Observable & Expt. value & Fit value \\ \hline
$\Delta m_{23}^2 / \Delta m_{12}^2$ & 20 -- 1670 & 376 \\
$\sin^2 2\theta_{12}$ & $>0.8$ & $0.88$ \\
$\sin^2 2\theta_{23}$ & $>0.8$ & $0.83$ \\
$\sin^2 2\theta_{13}$ & $<0.18$ & $0.10$\\
\hline
\end{tabular}}
\end{table}

The explosion of neutrino data and better bounds on rare FCNC
processes will provide ever more precise clues as to the nature of
flavor physics.  We have seen how well a rather small discrete gauge
group can accommodate all of the currently measured observables and
anticipate the refinement of such models in the future.

\nonumsection{Acknowledgments}
\noindent
A.A. and C.D.C. thank for support the National Science Foundation
under Grant No. PHY-9900657, and the Jeffress Memorial Trust under
Grant No.\ J-532.  RFL thanks the Department of Energy for support
under Grant No.\ DE-AC05-84ER40150 and the U. Maryland TQHN Group for
their hospitality.

\nonumsection{References}

\end{document}